\documentclass[twoside,english,5p, compress,url=false, sorted, times, numbers]{elsarticle}
\usepackage[T1]{fontenc}
\usepackage[latin9]{inputenc}
\usepackage{babel}
\usepackage{prettyref}
\usepackage{amsmath}
\usepackage{graphicx}
\usepackage{esint}
\usepackage[unicode=true,
 bookmarks=true,bookmarksnumbered=false,bookmarksopen=false,
 breaklinks=false,pdfborder={0 0 1},backref=false,colorlinks=false]
 {hyperref}

\makeatletter

\newcommand{\noun}[1]{\textsc{#1}}

\usepackage{standalone}

\usepackage[]{microtype}

\RequirePackage{colortbl, tabularx}
\@ifundefined{comment}{}
  {
   
  }%

\usepackage{prettyref}
\newrefformat{cha}{chapter~\ref{#1}}
\newrefformat{eq}{equation~\ref{#1}}

\newrefformat{sec}{section~\ref{#1}}
\newrefformat{sub}{subsection~\ref{#1}}
\newrefformat{fig}{figure~\ref{#1}}

\providecommand{\doi}[1]{%
  \begingroup
    \let\bibinfo\@secondoftwo
    \urlstyle{rm}%
    \href{http://dx.doi.org/#1}{%
      doi:\discretionary{}{}{}%
      \nolinkurl{#1}%
    }%
  \endgroup
}

\usepackage{enumitem}
\usepackage{amsmath,amssymb}
\usepackage{varwidth}

\providecommand{\doi}[1]{%
  \begingroup
    \let\bibinfo\@secondoftwo
    \urlstyle{rm}%
    \href{http://dx.doi.org/#1}{%
      doi:\discretionary{}{}{}%
      \nolinkurl{#1}%
    }%
  \endgroup
}
\cortext[cor1]{Corresponding author}

\@ifundefined{showcaptionsetup}{}{%
 \PassOptionsToPackage{caption=false}{subfig}}
\usepackage{subfig}
\makeatother

\begin{document}

\title{A new unified approach for modeling hot rolling of steel Part~1:
Comparison of models for recrystallization}

\author[imet,szmf]{J.~Orend}

\ead{jan.orend@tu-clausthal.de}

\author[szmf]{F.~Hagemann}

\ead{f.hagemann@sz.szmf.de}

\author[szmf]{F.~Klose}

\ead{f.klose@sz.szmf.de}

\author[szmf]{B.~Maas}

\ead{b.maas@sz.szmf.de}

\author[imet]{H.~Palkowski\corref{cor1}}

\ead{heinz.palkowski@tu-clausthal.de}

\address[imet]{Institute for Metallurgy, Robert-Koch-Stra{\ss}e 42, 38678 Clausthal,
Germany }

\address[szmf]{Salzgitter Mannesmann Forschung GmbH, Eisenh{\"u}ttenstra{\ss}e 99, 38239
Salzgitter, Germany}
\begin{abstract}
Models for the microstructure evolution during hot rolling are reviewed.
The basic macroscopic phenomena related to recrystallization are summarized.
Constitutive models based on semi empirical equations are compared
to more sophisticated models based on cellular automata, vertex and
Monte-Carlo-Potts methods. The applicability of each kind of model
approach for online and offline process control in steel industry
is discussed. While constitutive models are still state-of-the-art
for online process control, mesoscale models with a spatial representation
of the microstructure can provide better predictive capabilities at
the cost of long computation times. To fill this gap a new approach
based on modeling the interaction of an ensemble of multiple grains
is outlined and first simulation results are presented. The proposed
approach allows the unified modeling of dynamic, static and metadynamic
recrystallization as well as grain growth.

\end{abstract}
\begin{keyword}
Recrystallization\sep Model \sep Simulation \sep Process Control
\sep Grain Growth \sep Hot Rolling \sep Microstructure
\end{keyword}
\maketitle

\section{Introduction}

The microstructure evolution during hot rolling is of great interest
for the industrial production of steel and has therefore been subject
of research in the past decades. Recrystallization during and after
rolling is one of the mechanisms that can be used for grain refinement
\citep{sakai_dynamic_2014}. The resulting microstructure after rolling
is also important for the following phase transformation during cooling
having a significant influence on the mechanical properties of hot
rolled products. In consequence, controlling the recrystallization
is one opportunity for controlling the mechanical properties and in
turn gives room for saving expensive alloying elements. Therefore
models with good predictive capabilities are necessary.

Since dynamic recrystallization had been observed by \citet{greenwood_types_1939},
many theories and models for the description of recrystallization
during and after deformation have been developed. These models differ
in terms of complexity, characteristic length scale, practical usability
and the considered materials. Some of them have been reviewed by \citet{senuma_mathematical_1992},
\citet{militzer_computer_2007} and more recently by \citet{hallberg_approaches_2011}
and \citet{xiao_progress_2012}.

Additionally the usability of the different model types for technical
applications, like process control or numerical simulations of hot
forming processes, is discussed. One important aspect is the required
experimental effort necessary for determining material parameters.
Constitutive models use a large number of empirical parameters whereas
models with long computation times are not applicable for inverse
analysis methods which are often necessary for parameter identification
in an industrial context. Therefore a new approach that combines benefits
from constitutive and more sophisticated physical models is proposed.

\section{\label{sec:Phenomenons-of-Recrystallization}Phenomena related to
recrystallization}

\begin{figure}[tbph]
\includegraphics{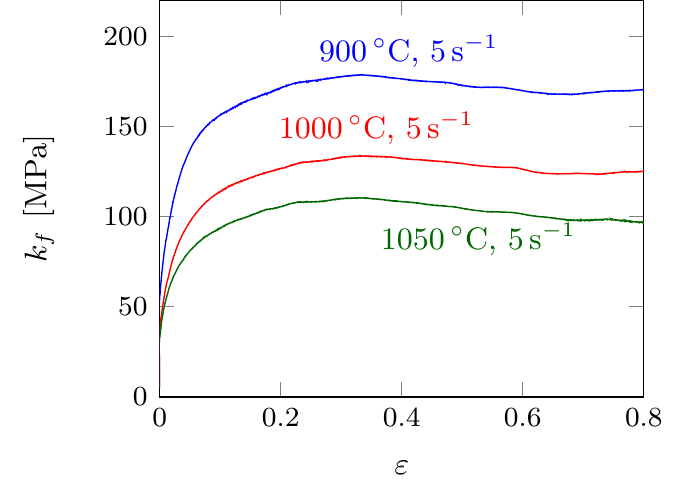}

\begin{centering}

\par\end{centering}

\caption{\label{fig:steady state}Flow stress curves at different temperatures
measured in cylindrical hot compression tests of 42CrMo4 steel samples.
After exceeding a specific plastic strain the flow stress decreases
due to dynamic recrystallization. }
\end{figure}

\begin{figure}[tbph]
\subfloat[$900\,^{\circ}\text{C}$, $5\,\mathrm{s^{-1}}$]{\begin{centering}
\includegraphics[width=0.45\columnwidth]{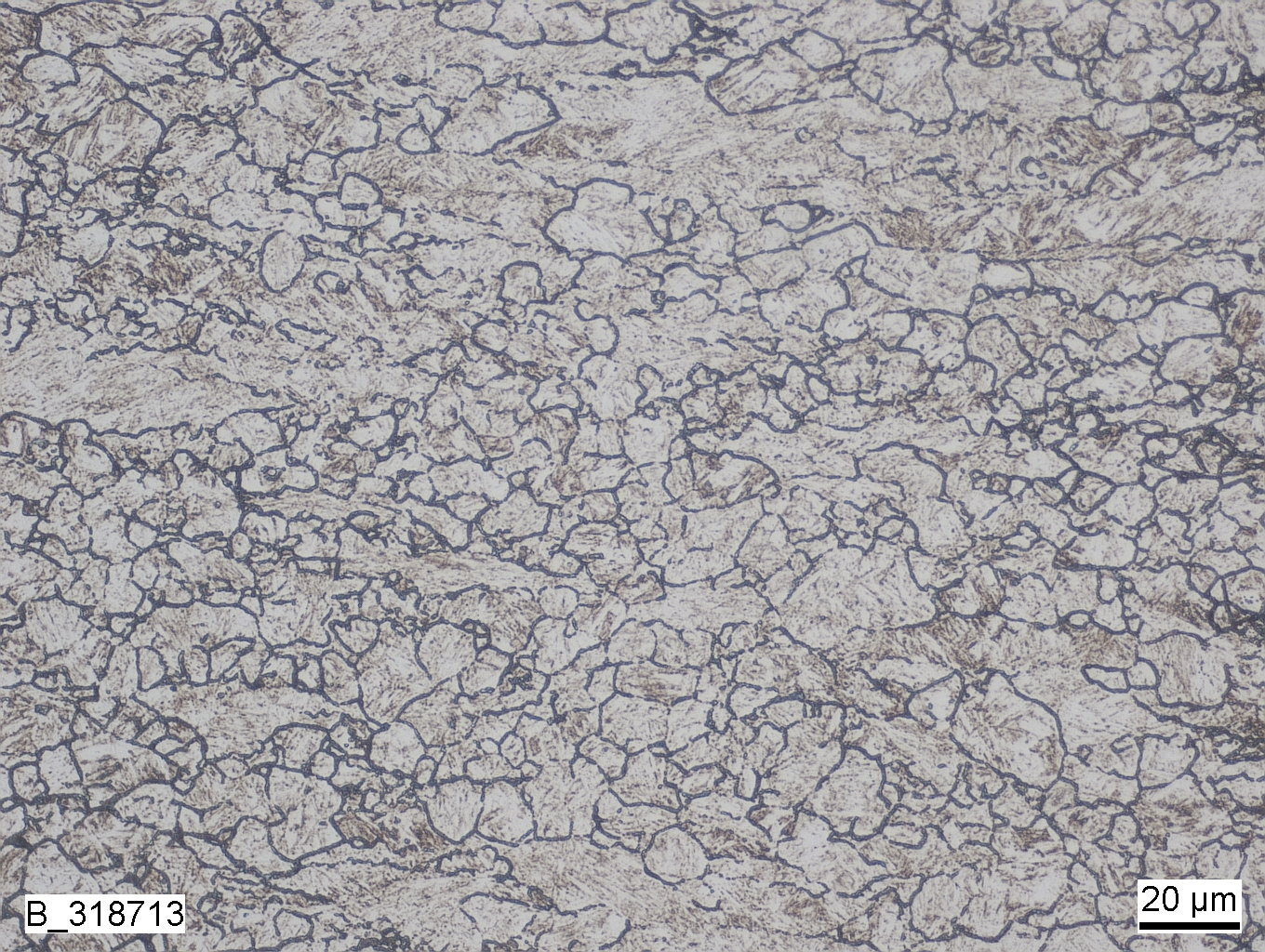}
\par\end{centering}

}\subfloat[$1000\,^{\circ}\text{C}$, $5\,\mathrm{s^{-1}}$]{\begin{centering}
\includegraphics[width=0.45\columnwidth]{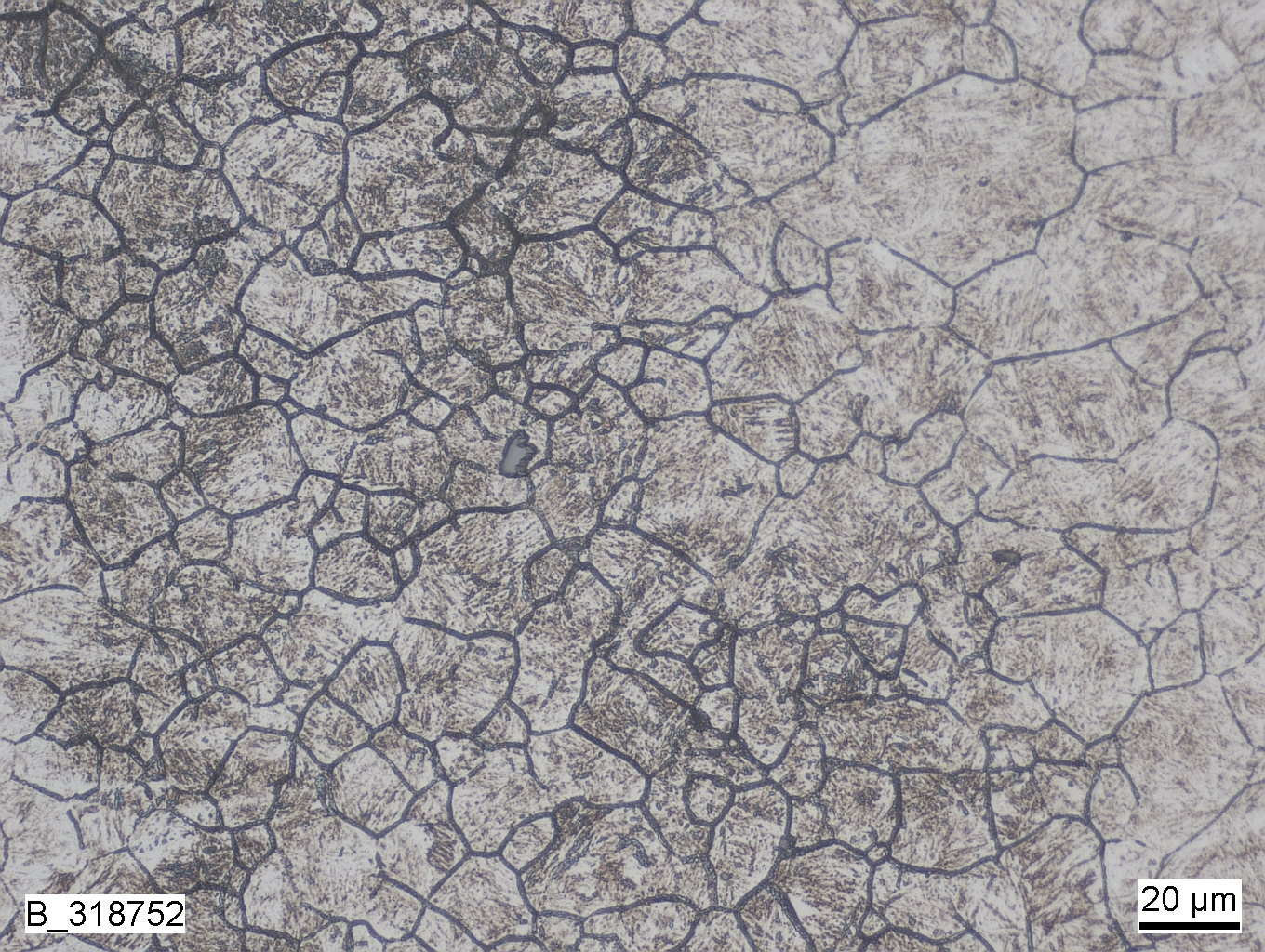}
\par\end{centering}

}

\caption{\label{fig:Revealed-prior-austenite}Revealed prior austenite grain
boundaries after hot deformation and quenching of 42CrMo4 to $\varepsilon=0.5$
with a strain rate of $\dot{\varepsilon}=5\mbox{\,\ensuremath{\mathrm{s^{-1}}}}$. }
\end{figure}

Several phenomena on macroscopic and microscopic scale can be related
to recrystallization. Some of them are summarized below.
\begin{itemize}
\item During deformation recrystallization can occur after exceeding a critical
strain referred to as dynamic recrystallization (\emph{DRX}) \citep{luton_dynamic_1969}.
In \prettyref{fig:steady state} flow curves from cylindrical hot
compression tests with 42CrMo4 steel are shown. For strains larger
than the peak strain a decrease of the flow stress can be observed
that is considered to be caused by \emph{DRX}. 
\item At high strain rates the flow stress monotonously tends to a lower
steady state stress after reaching the peak stress. For lower strain
rates a damped oscillation of the flow stress can be observed instead
what is referred to as cyclic recrystallization \citep{luton_dynamic_1969}. 
\item The strain $\varepsilon_{\mathrm{p}}$ at the peak stress increases
with strain rate and decreases with temperature \citep{luton_dynamic_1969}. 
\item The criteria for the transition from continuous to cyclic recrystallization
resulting in grain coarsening depends on the Zener-Holomon-parameter
and the initial grain size \citep{sakai_overview_1984}. 
\item The grain size during deformation in the steady state regime correlates
with the steady state stress as shown by \citet{sakai_dynamic_2014}. 
\item Recrystallization after deformation is called static recrystallization
(\emph{SRX}). If \emph{DRX} has occurred during deformation the recrystallization
kinetics strongly depends on the strain rate of the previous deformation.
This case is often referred to as metadynamic recrystallization (\emph{MDRX}).

\item After rapid changes of strain rate or temperature during deformation
the flow stress settles on a new steady state value. Before steady
state is reached transient oscillations can be observed \citep{sakai_dynamic_2014,frommert_dynamische_2008,frommert_mechanical_2009}.
\item If \emph{DRX} is completed, the austenite grain size after deformation
depends on temperature. In \prettyref{fig:Revealed-prior-austenite}
this is shown for two specimens of 42CrMo4 that have been deformed
to the logarithmic strain $\varepsilon=0.8$ with a strain rate of
$5\,\mathrm{s^{-1}}$ at temperatures of $900\,\mathrm{^{\circ}C}$
and $1000\,\mathrm{^{\circ}C}$. The specimens have been quenched
directly after deformation.
\end{itemize}

\section{Models for recrystallization}

One way to model the microstructure evolution is to model the observed
phenomena summarized above with a set of suitable mathematical functions
as proposed by \citet{sellars_recrystallization_1979}. These functions
include several material parameters that are derived from experimental
data. Another way is to model the physical mechanism like work hardening,
grain boundary movement and nucleation of new grains that lead to
the observed phenomena. To do so it is necessary to have some kind
of representation of the state of the microstructure. This can be
a set of scalar variables, a set of distribution functions or a spatial
representation of the microstructure. Such models are described in
\prettyref{sub:Inner-State-Variable} and \ref{sub:Models-with-spatial}.
In this paper we focus on model approaches that play a major role
in the development of microstructure models for practical applications.
Machine learning techniques like artificial neural networks and fuzzy
logic can be used for the prediction of the microstructure \citep{wang_artificial_2007,phaniraj_applicability_2003,mandal_modeling_2007}.
However, these methodologies provide ``Black-Box'' models that do
not allow for an insight view of the processes and are therefore not
discussed in this paper. 

In the following major approaches for modeling recrystallization are
described. At the end of each section the features and constraints
of each model are pointed out.

\subsection{\label{sub:Stateless-Models}Constitutive models}

Constitutive models provide a simple way of describing the microstructure
evolution mostly by closed form equation, which makes it possible
to use them even in simple spread sheet applications.

\subsubsection{\label{sub:Classical-models-for}Classical models for process control
applications}

\citet{sellars_recrystallization_1979} proposed a constitutive model
designed for the application of the simulation of the microstructure
evolution during hot rolling that built the starting point for many
other models developed by several groups \citep{ji_reduction_2006,gomez_modelling_2002,gomez_evolution_2009,sun_comparison_1997,davenport_development_2000,uranga_transition_2003,senuma_mathematical_1992,kim_modeling_2003}.
In this paper we only want to give a brief outline of the basic concept
behind this type of model. It has to be noted that a multitude of
modified or extended versions of the empirical submodels below exist
in literature. As input parameter these models use temperature $T$,
strain $\varepsilon$ and strain rate $\dot{\varepsilon}$ of each
deformation step, the initial average grain size $d_{0}$ and the
time $t$ after the prior deformation. The output parameters are the
recrystallized volume fraction $X$ and the average grain size $d$.
Constitutive microstructure evolution models distinguish between different
recrystallization phenomena (\emph{DRX}, \emph{MDRX}, \emph{SRX} and
grain growth) that are modeled separately. Each model contains material
parameters to be derived from experiments. The recrystallized volume
fraction and the prior austenite grain size have to be determined
after thermomechanical treatments under various processing conditions,
to allow for the determination of material parameters by inverse analysis
methods. Many models use the parameter 
\begin{equation}
Z=\dot{\varepsilon}\exp\left(\frac{Q_{\mathrm{def}}}{RT}\right)\,\mathrm{}\label{eq:Z}
\end{equation}
introduced by \citet{zener_effect_1944} with $Q_{\mathrm{def}}$
as an activation energy and $R=8.314\,\mathrm{\frac{J}{mol\, K}}$
as the universal gas constant to combine the dependency of strain
rate $\dot{\varepsilon}$ and temperature $T$ into one parameter.
The activation energy can be determined with the method proposed by
\citep{kugler_estimation_2004}. The recrystallized volume fraction
is often described by a modified version of the \noun{Johnson-Mehl-Avrami-Kolmogorov}
equation \citep{avrami_kinetics_1939}. For \emph{SRX} it can be written
as

\begin{equation}
X=1-\exp\left(-0.693\left(\frac{t}{t_{0.5}}\right)^{n}\right)\label{eq:Avrami}
\end{equation}
with $t_{0.5}$ being the time elapsed until the volume has recrystallized.
The time $t_{0.5}$ is calculated with an empirical equation of the
common form 
\begin{equation}
t_{0.5}=A_{\mathrm{SRX}}\varepsilon^{p_{\mathrm{SRX}}}d_{\mathrm{0}}^{q}Z^{r_{\mathrm{SRX}}}\exp\left(\frac{Q_{\mathrm{SRX}}}{RT}\right)\label{eq:t05srx}
\end{equation}
with $n,\, A_{\mathrm{SRX}},\, p_{\mathrm{SRX}},\, q,\, r_{\mathrm{SRX}}$
and $Q_{\mathrm{SRX}}$ as material parameters  and $d_{\mathrm{0}}$
as the average grain size prior to the previous deformation. The equation
can also be extended to take the effect of niobium into account as
proposed by \citet{hodgson_mathematical_1992}. The average diameter
of the recrystallized grains is then calculated with 
\begin{equation}
d_{\mathrm{SRX}}=\varepsilon^{p_{\mathrm{d}}}d_{\mathrm{0}}^{q_{\mathrm{d}}}\exp\left(\frac{Q_{\mathrm{D}}}{RT}\right)\,\label{eq:dsrx}
\end{equation}
where $p_{\mathrm{d}}\,,q_{\mathrm{d}}$ and $Q_{\mathrm{D}}$ are
material parameters. 

The onset of \emph{DRX} during deformation is described by a critical
strain 
\begin{equation}
\varepsilon_{\mathrm{c}}=A_{\mathrm{c}}d_{\mathrm{0}}^{p_{\mathrm{c}}}Z^{n_{\mathrm{c}}}\label{eq:emp_strain}
\end{equation}
that has to be exceeded with $A_{\mathrm{c}},\, p_{\mathrm{c}}$ and
$n_{\mathrm{c}}$ as material parameter. For the volume fraction recrystallized
by \emph{DRX} the time is replaced by strain resulting in

\begin{equation}
X_{\mathrm{DRX}}=1-\exp\left[B\left(\frac{\varepsilon-\varepsilon_{\mathrm{c}}}{\varepsilon_{\mathrm{X}}}\right)^{k}\right]\label{eq:X_drx}
\end{equation}
with $\varepsilon_{\mathrm{X}}$ being the strain at which the volume
fraction $\mathrm{X}$ has been recrystallized. The strain $\varepsilon_{\mathrm{X}}$
can be described by an empirical equation of the same form as \prettyref{eq:emp_strain}.
$B$ and $k$ are material constants. The grain size of the recrystallized
grains can be calculated using 
\begin{equation}
d_{\mathrm{DRX}}=CZ^{r}\,\label{eq:grain_semiemp}
\end{equation}
where $r$ and $C$ are material parameters. If \emph{DRX} has occurred
the kinetics of recrystallization after deformation depends on the
strain rate, i.e. metadynamic recrystallization. The time $t_{\mathrm{0.5}}$
decreases with increasing strain rate \citep{hodgson_mathematical_1992}.
Taking this effect into account \citeauthor{hodgson_mathematical_1992}
give $t_{0.5,\mathrm{MD}}$ as
\begin{equation}
t_{0.5,\mathrm{MD}}=k_{\mathrm{MD}}Z^{n_{\mathrm{MD}}}\exp\left(\frac{Q_{\mathrm{MD}}}{RT}\right)\,\label{eq:t05mdrx}
\end{equation}
with $k_{\mathrm{MD}}$, $n_{\mathrm{MD}}$ and $Q_{\mathrm{MD}}$
as material parameters \citep{hodgson_mathematical_1992}. The recrystallized
grain size by \emph{MDRX} can be calculated with an equation of the
form of \prettyref{eq:grain_semiemp} with an extra set of material
parameters for \emph{MDRX} $\left(C_{\text{\ensuremath{\mathrm{MDRX}}}},r_{\mathrm{MDRX}}\right)$.
For multistage deformation the effect of work hardening of the previous
deformations and of recrystallization on the dislocation density has
to be taken into account. To do so the concept of accumulated strain
is used by many authors \citep{ji_reduction_2006,sun_comparison_1997,hodgson_mathematical_1992}.
The strain of each pass is accumulated to $\varepsilon_{\mathrm{ACC}}$.
This accumulated strain is decreased depending on the recrystallized
volume fraction $X$ resulting in an effective strain $\varepsilon_{\mathrm{eff},i}$
that can be used for the calculation of the flow stress. For the calculation
of $\varepsilon_{\mathrm{eff},i}$ various approaches have been discussed
by \citet{hodgson_mathematical_1992}. One of them is a linear law
of mixtures as 
\begin{equation}
\varepsilon_{\mathrm{eff},i}=\varepsilon_{i}+\lambda\left(1-X\right)\varepsilon_{i-1}
\end{equation}
with $\lambda=1$ for C-Mn steels.

Since only average values are considered it is not possible to track
multiple volume fractions that have been recrystallized at different
time steps. The large number of material parameters make such models
easily adjustable to experimental data but also makes them demanding
on the scope of the experiments. Although such models cannot describe
the influence of the previous processing path due to the lack of inner
state variables, they show good agreement with experimental results
at steady state conditions, whereas cyclic recrystallization and transient
oscillations cannot be described.

\subsubsection{\label{sub:Luton-and-Sellars}Luton and Sellars model for cyclic
recrystallization}

While the models above aim for practical applications the following
model was a starting point for later models considering cyclic recrystallization.
Observing \emph{DRX} during hot torsion tests with nickel and nickel-iron
alloys \citet{luton_dynamic_1969} propose a model that describes
the transition from cyclic to steady state behavior. They assumed
that \emph{DRX} is initiated after exceeding a critical strain $\varepsilon_{\mathrm{c}}$
and introduced a strain $\varepsilon_{X}$ to characterize the strain
related to the time $t_{X}$ needed for a large fraction $X$ to recrystallize
under constant strain rate conditions. If $\varepsilon_{X}<\varepsilon_{\mathrm{c}}$
they suppose cyclic recrystallization and otherwise continuous recrystallization
behavior. 

They calculate the recrystallized volume fraction $X$ with 
\begin{equation}
X=1-\exp\left(-kt^{n}\right)
\end{equation}
where $t$ is the time after initiation of \emph{DRX} and $k$ and
$n$ are constants.

The flow stress is calculated as superposition of the flow stress
$\sigma_{\mathrm{e}}(\varepsilon)$ and $\sigma_{i}(\varepsilon)$
for each time increment $i$. The function $\sigma_{\mathrm{e}}(\varepsilon)$
describes the flow stress without recrystallization at the considered
constant temperature and strain rate. For each time increment $i$
a new function $\sigma_{i}\left(\varepsilon\right)=\sigma_{\mathrm{e}}\left(\varepsilon-\varepsilon_{i}\right)$
is introduced. The flow stress is then given as a superposition by
\begin{equation}
\sigma=\sum_{0}^{i}X_{i}\sigma_{i}+\left(1-\sum_{0}^{i}X_{i}\right)\sigma_{\mathrm{e}}\,
\end{equation}
where $X_{\mathrm{i}}$ is the volume fraction that has been recrystallized
in the increment $i$. 

This model is able to describe the transition from periodic to continuous
recrystallization. However, it does not describe the damping of the
oscillation for cyclic recrystallization.

\subsection{\label{sub:Inner-State-Variable}Inner state variable models}

Since microstructure evolution depends on the current microstructure,
models using inner state variables have been introduced. In this section
some early model approaches are discussed that were also the basis
for more recent models using spatial representations of the microstructure.

\subsubsection{\label{sub:St=0000FCwe-and-Ortner}St?we and Ortner}

\citet{stuwe_recrystallization_1974} suggest a model for \emph{DRX}
that describes the recrystallized volume fraction and the resulting
flow stress during deformation depending on strain rate and temperature
based on the evolution of the dislocation density. In this model the
average dislocation density can be seen as an inner state variable.
They calculate the flow stress $\sigma$ with
\begin{equation}
\sigma=AGb\sqrt{\overline{\rho}}\label{eq:taylor}
\end{equation}
with $b$ as the length of the burgers vector, $G$ as shear modulus,
$\overline{\rho}$ as average dislocation density and $A$ as a material
parameter.

The increase of the dislocation density during deformation due to
work hardening is 

\begin{equation}
\frac{\partial\rho}{\partial\varepsilon}=\frac{2\left(l+c\right)}{lcb}
\end{equation}
where $l$ and $c$ are the dimensions of the area swept by one dislocation
loop of the burgers vector $b$. In contrast to \citet{luton_dynamic_1969}
\citet{stuwe_recrystallization_1974} assume that \emph{DRX} starts
after the time 
\begin{equation}
t=t_{0}=\frac{\rho_{\mathrm{n}}}{\dot{\rho}}\label{eq:critical time}
\end{equation}
for a constant $\dot{\rho}$ after a critical dislocation density
$\rho_{\mathrm{n}}$ has been exceeded instead of using a critical
strain criterion. For spherical growth with a constant grain boundary
velocity $v_{\mathrm{n}}$ the recrystallized volume fraction $X$
after the time $t_{\mathrm{n}}$ is given by 
\begin{equation}
X=\left(\frac{v_{\mathrm{n}}\left(t_{\mathrm{n}}-t_{\mathrm{0}}\right)}{R}\right)^{3}\label{eq:stuewe_X}
\end{equation}
where $R$ is the traveling distance and for the given assumptions
the radius of the recrystallized grains. It has to be noted that $v_{n}$
depends on the grain boundary mobility that is a function of the temperature.
The average dislocation density $\overline{\rho}$ needed for the
calculation of the flow stress by \prettyref{eq:taylor} is determined
by integrating over the radius of the grains with 
\begin{equation}
\overline{\rho}=\frac{3}{4\pi R^{3}}\int_{0}^{R}\rho\left(r\right)4\pi r^{3}dr
\end{equation}
to take into account that the volume behind a moving boundary has
a very low dislocation density producing a gradient of $\rho$ from
the inner to the outer region of the grain. 

While the model can describe a transition from single peak to multiple
peak behavior for the combination of low strain rates and high temperatures,
the theoretical stress strain curves published by \citeauthor{stuwe_recrystallization_1974}
show a saw like shape that has not been observed in experiments. In
contrast to the constitutive models described above recrystallization
and the calculation of flow stress are combined into one model. The
model also relies on the assumption of constant strain rate and temperature.
Hence the model does not include the evolution of the grain size,
its effect on nucleation rate and grain boundary movement are not
considered. This issue has been addressed with the model developed
by Sandstr�m and Lagneborg described below.

\subsubsection{\label{sub:Sandstr=0000F6m-and-Lagneborg}Sandstr�m and Lagneborg}

In response to the models by \citeauthor{luton_dynamic_1969} and
\citeauthor{stuwe_recrystallization_1974}, \citet{sandstrom_model_1975-2}
point out, that the current grain size should also be taken into account.
They proposed a new model that describes the evolution of the dislocation
density distribution. They distinguish between the dislocation density
in the subgrain walls $\rho_{\mathrm{d}}$ and the homogeneous dislocation
density between the subgrain walls $\rho$. The volume distribution
of both kinds of dislocations are described by the functions $g\left(\rho,t\right)$
and $G\left(\rho_{\mathrm{d}},t\right)$ with 
\begin{eqnarray}
\int g\left(\rho,t\right)\mathrm{d}\rho & = & 1\\
\int G\left(\rho_{\mathrm{d}},t\right)\mathrm{d}\rho_{\mathrm{d}} & = & 1.
\end{eqnarray}
The time derivative of $\rho$ is given by

\begin{equation}
\frac{\mathrm{d}\rho}{\mathrm{d}t}=\frac{\dot{\varepsilon}}{bl}-2M\tau\rho^{2}
\end{equation}
considering work hardening and recovery. Here $l$ is the mean free
path of the dislocation and $M$ the dislocation mobility. Recovery
for the dislocations in the subgrain walls is neglected so that their
evolution is given by 
\begin{equation}
\frac{\mathrm{d}\rho_{\mathrm{d}}}{\mathrm{d}t}=\frac{\dot{\varepsilon}}{bl_{\mathrm{d}}}\label{eq:dis_subwalls}
\end{equation}
where the mean free path $l_{\mathrm{d}}$ is directly related to
the subgrain size. They conclude, that $l$ has to be much larger
than $l_{\mathrm{d}}$ if the main part of the dislocations are accumulated
in the subgrain walls which is reported by \citet{stuwe_recrystallization_1974}
to be also necessary for the calculation of realistic values of the
flow stress. According to \prettyref{eq:dis_subwalls} the relation
between the critical strain $\varepsilon_{\mathrm{cr}}$ and the critical
dislocation density $\rho_{\mathrm{cr}}$ for the onset of recrystallization
is given as 
\begin{equation}
\varepsilon_{\mathrm{cr}}=bl_{\mathrm{d}}\rho_{\mathrm{cr}}.
\end{equation}
The time derivative of the recrystallized volume fraction is given
by 
\begin{equation}
\frac{\mathrm{d}X}{\mathrm{d}t}=\int_{\rho_{\mathrm{cr}}}^{\infty}\frac{\xi\gamma_{\mathrm{D}}}{D}v\left(\rho_{\mathrm{d}}\right)G\left(\rho_{\mathrm{d}},t\right)\mathrm{d}\rho_{\mathrm{d}}
\end{equation}
where $D$ is the grain diameter and $\xi$ a constant given as $\xi=3$.
The velocity of the grain boundary $v\left(\rho\right)$ is given
by
\begin{equation}
v\left(\rho\right)=m\tau\rho
\end{equation}
with the grain boundary mobility $m$ and $\tau$ as dislocation line
energy. The parameter $\gamma_{\mathrm{D}}$ is called mobile fraction
and describes the effect that some of the grain boundaries do not
move. In the model of \citeauthor{sandstrom_model_1975-2} $\gamma_{\mathrm{D}}$
is assumed to be constant. 

The average stress is assumed to depend on the dislocation density
in the subgrains and is also calculated using \prettyref{eq:taylor}
but with the average dislocation density being 
\begin{equation}
\overline{\rho}=\int_{\rho_{0}}^{\rho_{\mathrm{s}}}\rho g\left(\rho,t\right)\mathrm{d\rho}
\end{equation}
where

\begin{equation}
\rho_{\mathrm{s}}=\sqrt{\frac{\dot{\varepsilon}}{2blM\tau}}
\end{equation}
is the dislocation density when work hardening and recovery are in
balance. While the dislocation density is described by distribution
functions the grain size is only represented by an average grain diameter
$D.$ The evolution of $D$ is described as 
\begin{equation}
\frac{\mathrm{d}D}{\mathrm{d}t}=m\frac{\gamma}{D}-D\frac{\mathrm{d}X}{\mathrm{dt}}\ln N\left(D\right)
\end{equation}
with the grain boundary mobility $m$, the grain boundary energy $\gamma$
and the amount of grains $N$ nucleated per parent grain after one
recrystallization cycle. 

Like the model by St?we and Ortner the model proposed by \citet{sandstrom_model_1974}
describes the recrystallization combined with the evolution of the
dislocation density. Additionally it also considers the evolution
of the average grain size. It is able to describe the damped oscillations
in flow curves at low strain rates. In comparison with data from experiments
with nickel the calculated peak strain at high strain rates is too
high. This is explained by not taking into account the strain rate
dependency of the mean free path $l$ and the dislocation mobility
$m$.

\subsection{\label{sub:Models-with-spatial}Models with spatial representation}

Instead of describing the microstructure only with average values
some more sophisticated models use a 2D or 3D spatial representation
of the microstructure. This representation then evolves according
to a set of certain rules that represent the physical mechanisms.
The common principle here is the minimization of the free energy.
For this there are various methodologies that are not only used for
modeling recrystallization but for a broad range of applications.
As for the models above temperature and strain rate are the input
parameters but here both parameters can change with time. These models
output a virtual representation of the grain morphology and the dislocation
density that can be used for the calculation of the flow stress.

\subsubsection{\label{sub:Cellular-Automata}Cellular Automata}

The Cellular Automata (\emph{CA}) method is based on the work of \citet{ulam_adventures_1991}
and \citet{von_neumann_general_1951}. A \emph{CA} consists of a set
of cells that are commonly ordered in a lattice. Each cell has a state
described by a set of variables and is related to a defined set of
cells that are called neighborhood. The transition of one state to
another is defined by a set of rules that is applied at each evolution
step. These rules can only rely on the state of one cell and its neighborhood
at the previous evolution step. Starting with \citet{hesselbarth_simulation_1991}
many groups have developed models for recrystallization utilizing
the \emph{CA} method \citep{ding_coupled_2001,kroc_application_2002,kuc_modelling_2011,kugler_study_2006,lee_cellular_2010,qian_cellular_2004,won_lee_numerical_2010,yazdipour_microstructural_2008,liu_simulation_1996}. 

In particular \citet{kugler_modeling_2004} used a two-dimensional
\emph{CA} with a rectangular grid for the simulation of multistage
deformations. To use the \emph{CA} to describe a virtual microstructure
the state of one cell consists of four variables including one variable
for the dislocation density $\rho_{i}$ and the crystal orientation.
The evolution of the dislocation density is calculated by integration
of

\begin{equation}
\frac{\partial\rho_{i}}{\partial\gamma}=k_{1}\sqrt{\rho_{i}}-k_{2}\rho_{i}\label{eq:dis_evo}
\end{equation}
for each cell where $\gamma$ is the shear strain, $k_{1}$ a hardening
coefficient and $k_{2}$ a strain rate and temperature depend parameter
describing recovery. The flow stress is calculated with \prettyref{eq:taylor}
using 

\begin{equation}
\bar{\rho}=\frac{1}{n}\sum_{i=1}^{n}\rho_{i}
\end{equation}
where $n$ is the number of cells. Adjacent cells with the same orientation
belong to the same grain. At the grain boundary the transition from
one orientation to another occurs depending on the grain boundary
velocity calculated by 

\begin{equation}
v=m\triangle f
\end{equation}
with $\triangle f$ as the driving force per unit area that is a function
of the difference of the dislocation densities and the grain boundary
energy. The grain boundary mobility $m$ is given by 

\begin{equation}
m=\frac{b\delta D}{k_{\mathrm{B}}T}\exp\left(\frac{Q_{\mathrm{b}}}{RT}\right)
\end{equation}
taking the characteristic grain boundary thickness $\delta$ and the
boundary self-diffusion coefficient $D$ into account. $R$ is the
universal gas constant, $k_{\mathrm{B}}$ is the Boltzmann constant
and $Q_{\mathrm{b}}$ is an activation energy. New grains are created
at the grain boundary if a critical dislocation density $\rho_{\mathrm{cr}}$
is exceeded with a probability
\begin{equation}
P=k\dot{\varepsilon}\exp\left(-\frac{\omega}{T}\right)\triangle t
\end{equation}
where $k$ and $\omega$ are material parameters and $\triangle t$
the time increment of one simulation step. 

\citeauthor{kugler_modeling_2004} show that their model is able to
describe the transition from continuous to cyclic softening at a constant
temperature with decreasing strain rate and for constant strain rate
with increasing temperature. They also performed numerical experiments
with two deformation stages for the analysis of the simulated post
deformation recrystallization. They found that exceeding a certain
transition strain during the first deformation results in a transition
from weak to strong strain rate dependency of the recrystallization
kinetics. It can be concluded that a model of this kind can describe
the transition from \emph{SRX} to \emph{MDRX} without introducing
submodels or special case handling for each recrystallization process
as it is necessary for the constitutive models described earlier.

\subsubsection{\label{sub:Monte-Carlo-Potts}Monte Carlo Potts method}

The application of the Monte Carlo Potts algorithm for modeling grain
growth and recrystallization started with the work of \citet{sahni_kinetics_1983}.
Similar to \emph{CA} models a lattice is used as representation of
the microstructure. Each of the cells, also called Monte Carlo Units
(\emph{MCU}), has a state $Q$ that indicates the misorientation and
the affiliation of the cell to one specific grain. Neighboring cells
with different values of $Q$ define a grain boundary. A basic algorithm
for the simulation of grain growth has been described by \citet{zollner_computer_2004}.
A random \emph{MCU} with the current state $Q_{x}$ is temporarily
set to a state $Q_{y}$ with $Q_{y}\neq Q_{x}$. The difference of
the energy $\triangle E$ between the states $Q_{x}$ and $Q_{y}$
is than calculated by a Hamiltonian $H$ that depends on the possible
misorientation between the cell and its neighbor cells. The resulting
energy difference $\triangle E$ is used to calculate a probability 

\begin{equation}
p=\begin{cases}
m, & \triangle E\leq0\\
m\exp\left(\frac{-\triangle E}{k_{\mathrm{B}}T}\right), & \triangle E>0
\end{cases}
\end{equation}
with $k_{\mathrm{B}}$ as Boltzmann constant and $m$ as grain boundary
mobility for the transition from the state $Q_{x}$ to $Q_{y}$. During
one Monte Carlo step one of these reorientation attempts is performed
for each cell. 

Hence this approach mimics the principle of least actions it can be
used to model the grain boundary movement without the definition of
specific evaluation rules, but to the cost of additional computation
time compared to \emph{CA} models.

\subsubsection{\label{sub:Vertex-Models}Vertex models}

\begin{figure}[tbph]
\begin{centering}
\includegraphics{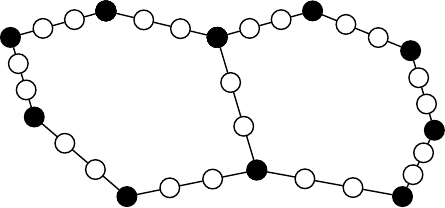}
\par\end{centering}

\caption{\label{fig:Illustration-of-the}Illustration of two grains in a simple
2D vertex model. Additionally to the vertices (filled) at the triple
points each grain boundary line is divided into segments by virtual
vertices.}
\end{figure}

Vertex models also use a spatial description of the microstructure.
In contrast to \emph{CA} models these models do not use a lattice.
Instead, the shape of the grains is only described by a set of geometrical
features like the vertices illustrated in \prettyref{fig:Illustration-of-the}.
This reduces the needed amount of memory and allows the simulation
of larger grain ensembles compared to the \emph{CA} method. The initialization
is often done by a Voronoi tesselation, sometimes followed by a simulated
annealing \citep{cocks_variational_1996,weygand_nucleation_2004},
that can be performed with the Lloyd-algorithm \citep{lloyd_least_1982}.

\citet{nagai_computer_1990} use a mesh with vertices at the triple
points of the grain boundaries in the 2D case. The grain boundaries
were modeled by straight lines connecting the vertices. They introduced
a force attached to the triple points resulting from the energy minimization
by optimizing the angles between the grain boundaries. This model
has been extended by several authors to allow curved boundaries. \citet{weygand_nucleation_2004}
insert additional vertices that divide the lines or planes representing
the grain boundary into multiple segments. \citeauthor{cocks_variational_1996}
\citep{cocks_variational_1996,gill_variational_1996} use cubic splines
for modeling curved boundaries in the two dimensional case. They also
show that the variational principle can be applied to derive equations
for the velocity of the vertices to model curvature-driven grain growth.
\citet{telley_laguerre_1996} and \citet{schule_justification_1996}
utilize the Laguerre tesselation, a Voronoi tesselation with radical
weighting, for the simulation of grain growth so only the coordinates
of the center and the weighting are needed to describe one grain. 

While the utilization of the vertex approach for the simulation of
grain growth is common, models are very rare that also consider nucleation.
Hence only the geometry of the grain boundaries is described this
approach does not provide a spatial dislocation density in the interior
of the grains like models based on the \emph{CA} approach or the Monte
Carlo Potts method. For modeling nucleation the dislocation density
has to be stored in an additional data structure.

\section{Ensemble model}

In the following the new model is outlined combining features of more
physical based approaches like \emph{CA} models mentioned above with
short computation times needed for process control applications. A
more detailed description of this model will follow in the second
part of this publication.

\subsection{Outline of a new approach}

\begin{figure}[tbph]
\begin{centering}
\includegraphics[width=1\columnwidth]{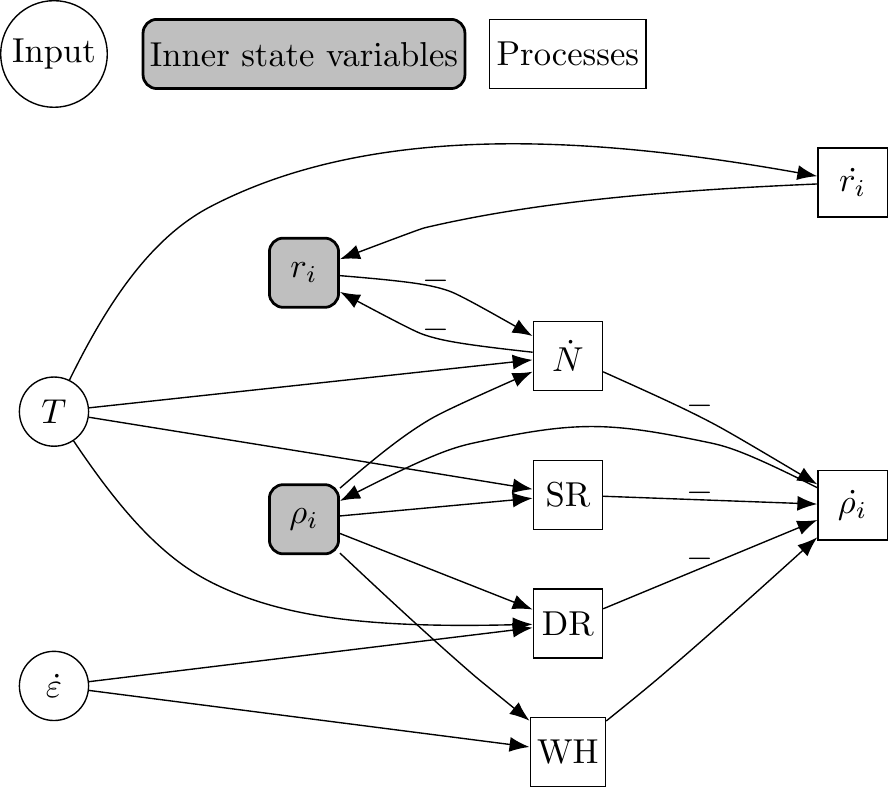}
\par\end{centering}

\caption{\label{fig:Graph-of-the}Graph of the processes considered for the
new model to be involved in \emph{DRX} and their dependencies. The
arrows indicate the direction of an influence. Arrows labeled with
(-) indicate a decrease of the influenced quantity. Strain rate $\dot{\varepsilon}$
and temperature $T$ are the input parameters. The state is defined
by a generalized variable $r_{i}$ describing the size and the dislocation
density $\rho_{i}$ of each grain. The state is changed by the processes
of work hardening (WH), dynamic recovery (DR) and static Recovery
(SR) effecting the time derivative of the dislocation density $\dot{\rho_{i}}$
and by nucleation of new grains ($\dot{N}$) and grain growth ($\dot{r_{i}}$). }
 
\end{figure}

The processes involved in recrystallization are time dependent and
influence each other as outlined in \prettyref{fig:Graph-of-the}.
In the model strain rate $\dot{\varepsilon}$ and temperature $T$
are the input parameters.
 Instead of using a spatial representation of the microstructure or
average values, the microstructure is described as an ensemble of
grains. Similar to a cell in a \emph{CA} model each grain $i$ has
state variables like the average dislocation density $\rho_{i}$ and
a generalized variable $r_{i}$ describing the grain size. These can
be evolved by the processes of nucleation of new grains ($\dot{N}$
in \prettyref{fig:Graph-of-the}), grain growth ($\dot{r_{i}}$),
dynamic recovery (DR), static recovery (SR) and work hardening (WH).
WH and SR are considered by the use of \prettyref{eq:dis_evo} for
the time derivative of the dislocation density $\dot{\rho}$. Each
grain has its own constant shape represented by a ratio 
\begin{equation}
\frac{V(r_{i})}{S(r_{i})}=\kappa_{i}r_{i}
\end{equation}
between grain volume $V\left(r_{i}\right)$ and surface $S\left(r_{i}\right)$
with $\kappa_{i}$ as a geometric constant that has to be set at the
initialization of the simulation. The ensemble consists of grains
in a specified volume that has to be kept constant. During the simulation
the size $r_{i}$ of each grain is adjusted by the integration of
$\dot{r_{i}}$ to reduce the free energy of the system in terms of
grain boundary energy and energy stored by dislocations in the volume
of the grain. During the grain boundary movement the power 
\begin{equation}
\dot{W}_{\mathrm{Diss}}=\frac{1}{2m}\dot{r_{i}}^{2}S\left(r_{i}\right)
\end{equation}
is dissipated with $m$ as grain boundary mobility. The time derivative
of $r_{i}$ is given by
\begin{equation}
\dot{r_{i}}=m\left(-3\kappa_{i}\left(\tau\rho_{i}+\lambda\right)-\frac{2}{r_{i}}\gamma\right)
\end{equation}
where $\tau$ is the dislocation line energy, $\gamma$ the grain
boundary energy per unit area and $\lambda$ a Lagrange multiplier.
Similar to the approach chosen by \citet{kugler_modeling_2004} new
grains are generated with a temperature dependent probability if a
critical dislocation density $\rho_{\mathrm{cr}}$ is exceeded. Preliminary
simulation results of this model are discussed below.

\subsection{Simulation results}

\begin{figure}[tbph]
\includegraphics{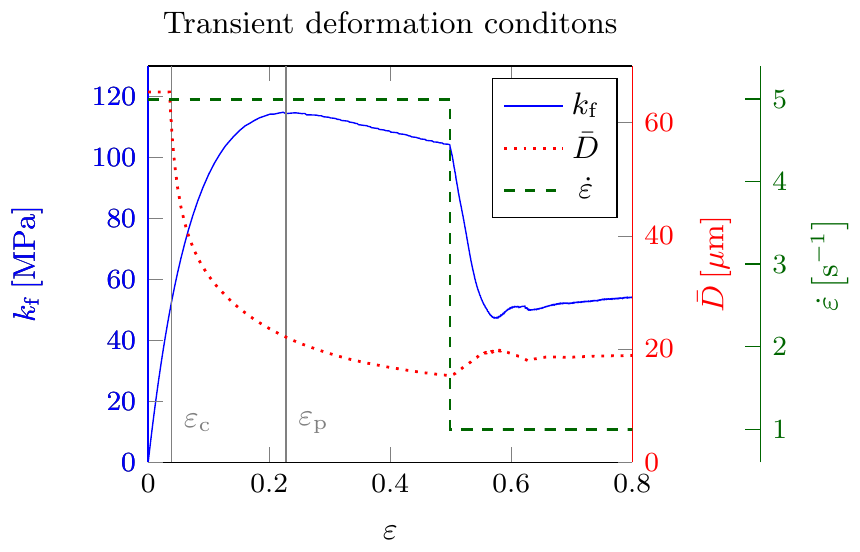}

\begin{centering}

\par\end{centering}

\caption{\label{fig:Results-from-simulated}Results from simulated deformation
under transient conditions. The calculated flow stress at a temperature
$T=1000\,\mathrm{^{\circ}C}$ for an initial average grain diameter
$d_{\mathrm{0}}=70\,\mathrm{\mu m}$ is shown. At the strain $\varepsilon=0.5$
the strain rate is changed from $5\,\mathrm{s^{-1}}$ to $1\,\mathrm{s^{-1}}$.
Following transient oscillation average grain size and flow stress
settle on new steady state values.}
\end{figure}

In order to investigate the dynamic behavior of the model simulations
under transient conditions were carried out. In \prettyref{fig:Results-from-simulated}
the results from a simulated deformation at a temperature of $1000\,\mathrm{^{\circ}C}$
with a rapid change of the strain rate form $5\,\mathrm{s^{-1}}$
to $1\,\mathrm{s^{-1}}$ at $\varepsilon=0.5$ are shown. The flow
stress first increases until it reaches a maximum at $\varepsilon_{\mathrm{p}}$.
During the first deformation stage a decrease of the average grain
size can be observed starting at a strain $\varepsilon_{\mathrm{c}}$
that is much smaller than $\varepsilon_{\mathrm{p}}$. After the strain
rate change to the smaller strain rate a transient oscillation of
the average grain diameter and the flow stress can be observed. This
behavior has also been observed in simulations with a \emph{CA} model
by \citet{kroc_application_2002} and in experiments by \citet{sakai_overview_1984}
and \citet{frommert_dynamische_2008}.

\begin{figure}[tbph]

\includegraphics{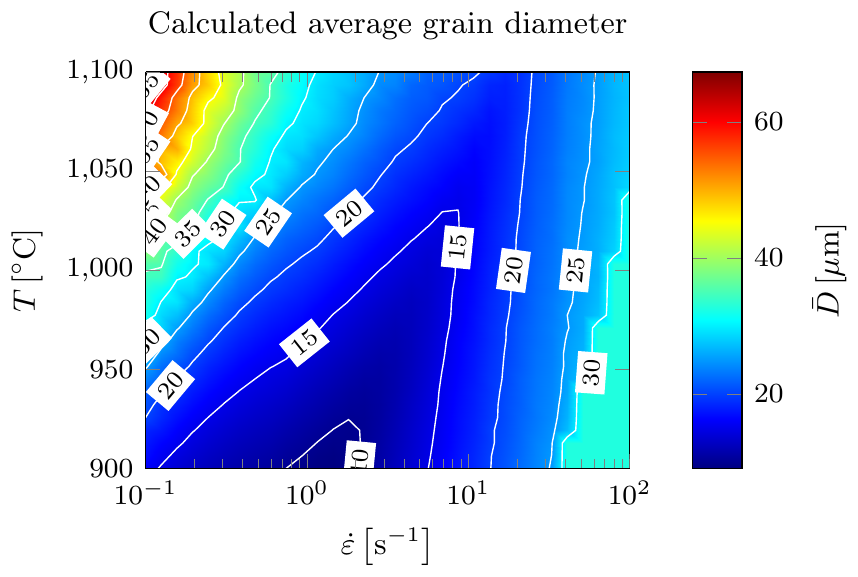}

\caption{\label{fig:Grain-size-map}Grain size map for one simulated deformation.
For a deformation with $\varepsilon=0.5$ and an initial grain diameter
of $d_{\mathrm{0}}=35\,\mathrm{\mu m}$ the calculated average value
of the grain diameter immediately after deformation for different
strain rates and temperatures is shown. For the combination of low
strain rates and high temperatures the model predicts grain coarsening,
at high strain rates recrystallization is not completed.}
\end{figure}

To illustrate the dependency on strain rate and temperature the average
grain size after a deformation step is shown in \prettyref{fig:Grain-size-map}.
For the combination of low strain rate and high temperature the model
predicts grain coarsening. In this case the nucleation rate is relatively
low in relation to the growth rate of the grains. At high strain rates
$\dot{\varepsilon}>20\,\mathrm{s^{-1}}$ the available time for recrystallization
decreases, so at low temperatures the grain size remains at the initial
state. At higher temperatures where recrystallization is faster there
is still grain refinement. The minimum of the grain size is at the
lowest temperature in a strain rate regime where $DRX$ is completed
and the strain rate is high enough to induce a high nucleation rate
by a high work hardening rate. 

\section{Discussion}

\begin{figure*}[t]
\begin{centering}
\includegraphics[width=1\textwidth]{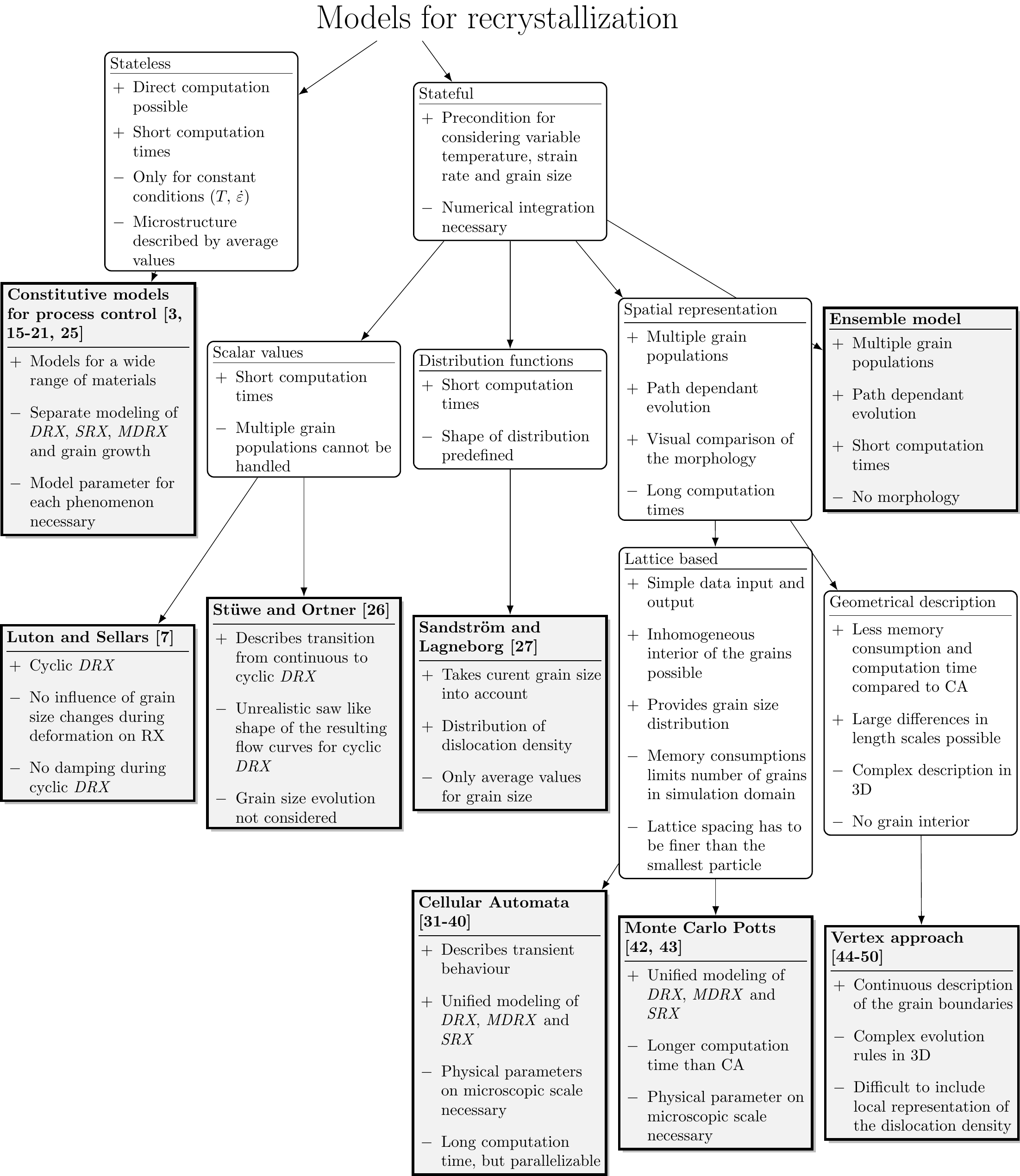}
\par\end{centering}

\caption{\label{fig:Comparison-of-different}Hierarchical overview of the features
and constraints of different model approaches for recrystallization.
The models (highlighted with a thick border) are divided in several
branches based on the underlying modeling techniques.}
\end{figure*}

The models described above diverge in their underlying assumptions
and modeling techniques. The major features and constraints are summarized
in \prettyref{fig:Comparison-of-different}.
 For the use of microstructure evolution models in technical applications,
like forming simulations or process control, a minimum of computational
effort is preferred. Constitutive models for the recrystallization
during hot deformation as described in \prettyref{sub:Stateless-Models}
support fast calculations of average values describing the microstructure
like grain diameter or recrystallized volume fraction. Due to their
simplicity and low computation times these kinds of models are popular
for process control applications. However, there are some drawbacks
for the practical application that give motivation for the development
of models that include more knowledge about the physical processes. 

Constitutive models describe the microstructure evolution in a phenomenological
way, so each phenomenon like flow stress, \emph{DRX}, \emph{SRX},
\emph{MDRX} and grain growth are described individually by empirical
submodels with their own sets of material parameters. To determine
these parameters for one steel grade each phenomenon has to be investigated
experimentally under a wide range of process conditions. However,
steel companies for example produce a wide range of different steel
grades, so there is a great interest in reducing the required experimental
work. In particular the determination of the grain size is difficult
since the prior austenite grain boundaries cannot easily be revealed
with metallographic techniques for a lot of modern low carbon steels. 

Another motivation for the usage of less phenomenological but more
physical based models is to increase the predictive capabilities.
Numerical simulation of an industrial hot rolling process using models
developed on the basis of laboratory tests are in most cases extrapolations
of the underlying experimental data. In a typical hot rolling process
strain rates up to $150\,\frac{1}{\mathrm{s}}$ and overall strains
up to $\varepsilon=5$ are possible. These conditions are difficult
to simulate experimentally on a laboratory scale. During multistage
deformation recrystallization may not always be completed. The microstrucuture
can consist of multiple generations of grains that have been recrystallized
at different points of time. This can not be modeled adequately when
only considering average values of the grain size and the recrystallized
volume fraction.

Mesoscale models that use a spatial representation of the microstructure
give an opportunity to simulate the physical processes involved in
recrystallization during hot rolling. In particular models based on
the Cellular Automata method show promising results like the simulation
of \emph{DRX} under transient conditions by \citet{kroc_application_2002}
that show similar patterns as the experimental results from \citet{frommert_dynamische_2008}
and \citet{sakai_dynamic_2014}. The simulation results of post-deformation
recrystallization by \citet{kugler_modeling_2004} show that\emph{
MDRX} and \emph{SRX} can be described with one unified model. This
can be explained by the influence of the current grain size on the
recrystallization kinetics. The size of the grains recrystallized
by \emph{DRX} depends on strain rate and temperature. Hence the kinetics
of \emph{MDRX} also depends on strain rate and temperature of the
previous deformation. However the initial grain size for \emph{SRX}
does not change due to the deformation condition because in this case
no \emph{DRX} is involved. So \emph{SRX} depends more on temperature
and the grain size prior to deformation than on strain rate. Other
benefits of such models are, that the virtual microstructure can easily
be compared with a real microstructure. It is also possible to track
multiple generation of recrystallized grains. However, the \emph{CA}
method is very computationally and memory intensive, especially if
a three-dimensional lattice is used. It is therefore not suited yet
for a direct usage in forming simulations or process control applications. 

The proposed ensemble model describes each grain with a set of scalar
variables. Compared to \emph{CA} models this reduces the required
amount of memory and computation cycles proportional to the number
of cells needed to describe one grain in a \emph{CA} lattice. A spherical
grain with a grain diameter of $20\,\mathrm{\mu m}$ fills approximately
4200 cells in a three-dimensional lattice with an edge length of $2\,\mathrm{\mu m}$
of each cell. The trade off is that the model only provides the distribution
of the grain size rather than a spatial representation of the microstructure.
The model holds no information about the relationship between the
grains. It is not possible to take into account the difference of
the misorientation between the grains for the calculations of the
grain boundary energy, so an average value is used. But like in vertex
models the grains are not discretized so grain boundary movement can
be described continuously. Similar to \emph{CA} models the ensemble
model can describe the transient oscillation that can be observed
at transient conditions during deformation. The system does not change
its state instantly to the configuration with the lowest free energy.
Instead, work has to be performed to move the grain boundaries. This
is a time dependent and dissipative process that follows the principle
of least actions. The path with the lowest energy dissipation is in
some cases an oscillation of the average grain size and the flow stress.

\section{Conclusion}

Mesoscale models allow the modeling of the microstructure evolution
during and after hot deformation by describing the underlying physical
processes. Phenomena modeled separately by constitutive models like
\emph{DRX}, \emph{SRX, MDRX} and grain growth can be described in
a unified manner. The outlined ensemble model makes some approaches
known from mesoscale models practically useable for through process
modeling and process control application by a massive reduction of
the required computational effort. 

All models rely on proper material parameters. Future work should
therefore focus on methods for the determination of model parameters,
especially for the models based on physical approaches. One way to
determine the parameters from experimental data is the use of the
inverse analysis method. This implies the execution of many simulation
cycles that are time consuming for complex models. The usage of the
ensemble model can be a solution to make such analysis feasible.\clearpage{}


\bibliographystyle{elsarticle-num-names}
\phantomsection\addcontentsline{toc}{section}{\refname}\bibliography{zotero}

\begin{thebibliography}{50}
\providecommand{\natexlab}[1]{#1}
\providecommand{\url}[1]{\texttt{#1}}
\providecommand{\urlprefix}{URL }
\expandafter\ifx\csname urlstyle\endcsname\relax
  \providecommand{\doi}[1]{doi:\discretionary{}{}{}#1}\else
  \providecommand{\doi}[1]{doi:\discretionary{}{}{}\begingroup
  \urlstyle{rm}\url{#1}\endgroup}\fi
\providecommand{\bibinfo}[2]{#2}

\bibitem[{Sakai et~al.(2014)Sakai, Belyakov, Kaibyshev, Miura, and
  Jonas}]{sakai_dynamic_2014}
\bibinfo{author}{T.~Sakai}, \bibinfo{author}{A.~Belyakov},
  \bibinfo{author}{R.~Kaibyshev}, \bibinfo{author}{H.~Miura},
  \bibinfo{author}{J.~J. Jonas}, \bibinfo{title}{Dynamic and post-dynamic
  recrystallization under hot, cold and severe plastic deformation conditions},
  \bibinfo{journal}{Prog. Mater Sci.} \bibinfo{volume}{60}
  (\bibinfo{year}{2014}) \bibinfo{pages}{130--207}, ISSN
  \bibinfo{issn}{0079-6425},
  \doi{\bibinfo{doi}{10.1016/j.pmatsci.2013.09.002}}.

\bibitem[{Greenwood and Worner(1939)}]{greenwood_types_1939}
\bibinfo{author}{J.~N. Greenwood}, \bibinfo{author}{H.~K. Worner},
  \bibinfo{title}{Types of creep curve obtained with lead and its dilute
  alloys}, \bibinfo{journal}{Journal Institute of Metals} \bibinfo{volume}{64}
  (\bibinfo{year}{1939}) \bibinfo{pages}{135}.

\bibitem[{Senuma et~al.(1992)Senuma, Suehiro, and
  Yada}]{senuma_mathematical_1992}
\bibinfo{author}{T.~Senuma}, \bibinfo{author}{M.~Suehiro},
  \bibinfo{author}{H.~Yada}, \bibinfo{title}{Mathematical models for predicting
  microstructural evolution and mechanical properties of hot strips},
  \bibinfo{journal}{{ISIJ} Int.} \bibinfo{volume}{32}~(\bibinfo{number}{3})
  (\bibinfo{year}{1992}) \bibinfo{pages}{423{\textendash}432}.

\bibitem[{Militzer(2007)}]{militzer_computer_2007}
\bibinfo{author}{M.~Militzer}, \bibinfo{title}{Computer simulation of
  microstructure evolution in low carbon sheet steels},
  \bibinfo{journal}{{ISIJ} international}
  \bibinfo{volume}{47}~(\bibinfo{number}{1}) (\bibinfo{year}{2007})
  \bibinfo{pages}{1{\textendash}15}.

\bibitem[{Hallberg(2011)}]{hallberg_approaches_2011}
\bibinfo{author}{H.~Hallberg}, \bibinfo{title}{Approaches to modeling of
  recrystallization}, \bibinfo{journal}{Metals}
  \bibinfo{volume}{1}~(\bibinfo{number}{1}) (\bibinfo{year}{2011})
  \bibinfo{pages}{16{\textendash}48}.

\bibitem[{Xiao et~al.(2012)Xiao, Chen, Li, and Li}]{xiao_progress_2012}
\bibinfo{author}{N.~Xiao}, \bibinfo{author}{Y.~Chen}, \bibinfo{author}{D.~Li},
  \bibinfo{author}{Y.~Li}, \bibinfo{title}{Progress in mesoscopic modeling of
  microstructure evolution in steels}, \bibinfo{journal}{Sci. China Technol.
  Sci.} \bibinfo{volume}{55}~(\bibinfo{number}{2}) (\bibinfo{year}{2012})
  \bibinfo{pages}{341--356}, ISSN \bibinfo{issn}{1674-7321, 1869-1900},
  \doi{\bibinfo{doi}{10.1007/s11431-011-4699-z}}.

\bibitem[{Luton and Sellars(1969)}]{luton_dynamic_1969}
\bibinfo{author}{M.~Luton}, \bibinfo{author}{C.~Sellars},
  \bibinfo{title}{Dynamic recrystallization in nickel and nickel-iron alloys
  during high temperature deformation}, \bibinfo{journal}{Acta Metall.}
  \bibinfo{volume}{17}~(\bibinfo{number}{8}) (\bibinfo{year}{1969})
  \bibinfo{pages}{1033--1043}, ISSN \bibinfo{issn}{0001-6160},
  \doi{\bibinfo{doi}{10.1016/0001-6160(69)90049-2}}, \bibinfo{note}{00398}.

\bibitem[{Sakai and Jonas(1984)}]{sakai_overview_1984}
\bibinfo{author}{T.~Sakai}, \bibinfo{author}{J.~J. Jonas},
  \bibinfo{title}{Overview no. 35 Dynamic recrystallization: Mechanical and
  microstructural considerations}, \bibinfo{journal}{Acta Metall.}
  \bibinfo{volume}{32}~(\bibinfo{number}{2}) (\bibinfo{year}{1984})
  \bibinfo{pages}{189{\textendash}209}.

\bibitem[{Frommert(2008)}]{frommert_dynamische_2008}
\bibinfo{author}{M.~M. Frommert}, \bibinfo{title}{Dynamische Rekristallisation
  unter konstanten und transienten Umformbedingungen},
  \bibinfo{publisher}{Cuvillier Verlag}, ISBN \bibinfo{isbn}{9783867274968},
  \bibinfo{note}{00001}, \bibinfo{year}{2008}.

\bibitem[{Frommert and Gottstein(2009)}]{frommert_mechanical_2009}
\bibinfo{author}{M.~Frommert}, \bibinfo{author}{G.~Gottstein},
  \bibinfo{title}{Mechanical behavior and microstructure evolution during
  steady-state dynamic recrystallization in the austenitic steel 800H},
  \bibinfo{journal}{Mater. Sci. Eng., A}
  \bibinfo{volume}{506}~(\bibinfo{number}{1{\textendash}2})
  (\bibinfo{year}{2009}) \bibinfo{pages}{101--110}, ISSN
  \bibinfo{issn}{0921-5093}, \doi{\bibinfo{doi}{10.1016/j.msea.2008.11.035}}.

\bibitem[{Sellars and Whiteman(1979)}]{sellars_recrystallization_1979}
\bibinfo{author}{C.~Sellars}, \bibinfo{author}{J.~Whiteman},
  \bibinfo{title}{Recrystallization and grain growth in hot rolling},
  \bibinfo{journal}{Met. Sci.} \bibinfo{volume}{3}~(\bibinfo{number}{4})
  (\bibinfo{year}{1979}) \bibinfo{pages}{187{\textendash}194}.

\bibitem[{Wang et~al.(2007)Wang, Lai, and Sun}]{wang_artificial_2007}
\bibinfo{author}{Q.~Wang}, \bibinfo{author}{J.~Lai}, \bibinfo{author}{D.~L.
  Sun}, \bibinfo{title}{Artificial neural network models for predicting flow
  stress and microstructure evolution of a hydrogenized titanium alloy},
  \bibinfo{journal}{Key Eng. Mater.} \bibinfo{volume}{353-358}
  (\bibinfo{year}{2007}) \bibinfo{pages}{541--544}, ISSN
  \bibinfo{issn}{1662-9795},
  \doi{\bibinfo{doi}{10.4028/www.scientific.net/KEM.353-358.541}}.

\bibitem[{Phaniraj and Lahiri(2003)}]{phaniraj_applicability_2003}
\bibinfo{author}{M.~P. Phaniraj}, \bibinfo{author}{A.~K. Lahiri},
  \bibinfo{title}{The applicability of neural network model to predict flow
  stress for carbon steels}, \bibinfo{journal}{J. Mater. Process. Technol.}
  \bibinfo{volume}{141}~(\bibinfo{number}{2}) (\bibinfo{year}{2003})
  \bibinfo{pages}{219--227}, ISSN \bibinfo{issn}{0924-0136},
  \doi{\bibinfo{doi}{10.1016/S0924-0136(02)01123-8}}.

\bibitem[{Mandal et~al.(2007)Mandal, Sivaprasad, and
  Dube}]{mandal_modeling_2007}
\bibinfo{author}{S.~Mandal}, \bibinfo{author}{P.~V. Sivaprasad},
  \bibinfo{author}{R.~K. Dube}, \bibinfo{title}{Modeling microstructural
  evolution during dynamic recrystallization of alloy D9 using artificial
  neural network}, \bibinfo{journal}{J. Mater. Eng. Perform.}
  \bibinfo{volume}{16}~(\bibinfo{number}{6}) (\bibinfo{year}{2007})
  \bibinfo{pages}{672--679}, ISSN \bibinfo{issn}{1059-9495, 1544-1024},
  \doi{\bibinfo{doi}{10.1007/s11665-007-9098-z}}.

\bibitem[{Ji and Shivpuri(2006)}]{ji_reduction_2006}
\bibinfo{author}{M.~Ji}, \bibinfo{author}{R.~Shivpuri},
  \bibinfo{title}{Reduction of random seams in hot rolling through {FEM} based
  sensitivity analysis}, \bibinfo{journal}{Mater. Sci. Eng., A}
  \bibinfo{volume}{425}~(\bibinfo{number}{1{\textendash}2})
  (\bibinfo{year}{2006}) \bibinfo{pages}{156--166}, ISSN
  \bibinfo{issn}{0921-5093}, \doi{\bibinfo{doi}{10.1016/j.msea.2006.03.071}}.

\bibitem[{G{\'o}mez and P{\'e}rez(2002)}]{gomez_modelling_2002}
\bibinfo{author}{G.~R. G{\'o}mez}, \bibinfo{author}{T.~P{\'e}rez},
  \bibinfo{title}{Modelling the microstructural evolution during hot rolling},
  \bibinfo{journal}{Latin American applied research}
  \bibinfo{volume}{32}~(\bibinfo{number}{3}) (\bibinfo{year}{2002})
  \bibinfo{pages}{253{\textendash}256}.

\bibitem[{G{\'o}mez et~al.(2009)G{\'o}mez, Rancel, Fern{\'a}ndez, and
  Medina}]{gomez_evolution_2009}
\bibinfo{author}{M.~G{\'o}mez}, \bibinfo{author}{L.~Rancel},
  \bibinfo{author}{B.~J. Fern{\'a}ndez}, \bibinfo{author}{S.~F. Medina},
  \bibinfo{title}{Evolution of austenite static recrystallization and grain
  size during hot rolling of a V-microalloyed steel}, \bibinfo{journal}{Mater.
  Sci. Eng., A} \bibinfo{volume}{501}~(\bibinfo{number}{1{\textendash}2})
  (\bibinfo{year}{2009}) \bibinfo{pages}{188--196}, ISSN
  \bibinfo{issn}{0921-5093}, \doi{\bibinfo{doi}{10.1016/j.msea.2008.09.074}}.

\bibitem[{Sun and Hawbolt(1997)}]{sun_comparison_1997}
\bibinfo{author}{W.~Sun}, \bibinfo{author}{E.~Hawbolt},
  \bibinfo{title}{Comparison between static and metadynamic
  recrystallization-an application to the hot rolling of steels},
  \bibinfo{journal}{{ISIJ} Int.} \bibinfo{volume}{37}~(\bibinfo{number}{10})
  (\bibinfo{year}{1997}) \bibinfo{pages}{1000{\textendash}1009}.

\bibitem[{Davenport et~al.(2000)Davenport, Silk, Sparks, and
  Sellars}]{davenport_development_2000}
\bibinfo{author}{S.~Davenport}, \bibinfo{author}{N.~Silk},
  \bibinfo{author}{C.~Sparks}, \bibinfo{author}{C.~Sellars},
  \bibinfo{title}{Development of constitutive equations for modelling of hot
  rolling}, \bibinfo{journal}{Mater. Sci. Technol.} \bibinfo{volume}{16}
  (\bibinfo{year}{2000}) \bibinfo{pages}{539{\textendash}546}.

\bibitem[{Uranga et~al.(2003)Uranga, Fern{\'a}ndez, L{\'o}pez, and
  Rodriguez-Ibabe}]{uranga_transition_2003}
\bibinfo{author}{P.~Uranga}, \bibinfo{author}{A.~Fern{\'a}ndez},
  \bibinfo{author}{B.~L{\'o}pez}, \bibinfo{author}{J.~Rodriguez-Ibabe},
  \bibinfo{title}{Transition between static and metadynamic recrystallization
  kinetics in coarse Nb microalloyed austenite}, \bibinfo{journal}{Materials
  Science and Engineering: A}
  \bibinfo{volume}{345}~(\bibinfo{number}{1{\textendash}2})
  (\bibinfo{year}{2003}) \bibinfo{pages}{319--327}, ISSN
  \bibinfo{issn}{0921-5093},
  \doi{\bibinfo{doi}{10.1016/S0921-5093(02)00510-5}}, \bibinfo{note}{00070}.

\bibitem[{Kim et~al.(2003)Kim, Lee, Lee, and Yoo}]{kim_modeling_2003}
\bibinfo{author}{S.-I. Kim}, \bibinfo{author}{Y.~Lee}, \bibinfo{author}{D.-L.
  Lee}, \bibinfo{author}{Y.-C. Yoo}, \bibinfo{title}{Modeling of {AGS} and
  recrystallized fraction of microalloyed medium carbon steel during hot
  deformation}, \bibinfo{journal}{Mater. Sci. Eng., A}
  \bibinfo{volume}{355}~(\bibinfo{number}{1{\textendash}2})
  (\bibinfo{year}{2003}) \bibinfo{pages}{384--393}, ISSN
  \bibinfo{issn}{0921-5093},
  \doi{\bibinfo{doi}{10.1016/S0921-5093(03)00104-7}}.

\bibitem[{Zener and Hollomon(1944)}]{zener_effect_1944}
\bibinfo{author}{C.~Zener}, \bibinfo{author}{J.~Hollomon},
  \bibinfo{title}{Effect of strain rate upon plastic flow of steel},
  \bibinfo{journal}{J. Appl. Phys.} \bibinfo{volume}{15} (\bibinfo{year}{1944})
  \bibinfo{pages}{22}, \bibinfo{note}{00000}.

\bibitem[{Kugler et~al.(2004)Kugler, Knap, Palkowski, and
  Turk}]{kugler_estimation_2004}
\bibinfo{author}{G.~Kugler}, \bibinfo{author}{M.~Knap},
  \bibinfo{author}{H.~Palkowski}, \bibinfo{author}{R.~Turk},
  \bibinfo{title}{Estimation of activation energy or Calculating the hot
  workability properties of metals}, \bibinfo{journal}{Metalurgija}
  \bibinfo{volume}{43}~(\bibinfo{number}{4}) (\bibinfo{year}{2004})
  \bibinfo{pages}{267{\textendash}272}.

\bibitem[{Avrami(1939)}]{avrami_kinetics_1939}
\bibinfo{author}{M.~Avrami}, \bibinfo{title}{Kinetics of phase change. I
  General theory}, \bibinfo{journal}{J. Chem. Phys.} \bibinfo{volume}{7}
  (\bibinfo{year}{1939}) \bibinfo{pages}{1103}.

\bibitem[{Hodgson and Gibbs(1992)}]{hodgson_mathematical_1992}
\bibinfo{author}{P.~D. Hodgson}, \bibinfo{author}{R.~K. Gibbs},
  \bibinfo{title}{A Mathematical model to predict the mechanical properties of
  hot rolled C-Mn and microalloyed steels}, \bibinfo{journal}{{ISIJ} Int.}
  \bibinfo{volume}{32}~(\bibinfo{number}{12}) (\bibinfo{year}{1992})
  \bibinfo{pages}{1329--1338}, ISSN \bibinfo{issn}{0915-1559},
  \doi{\bibinfo{doi}{10.2355/isijinternational.32.1329}}.

\bibitem[{St{\"u}we and Ortner(1974)}]{stuwe_recrystallization_1974}
\bibinfo{author}{H.~P. St{\"u}we}, \bibinfo{author}{B.~Ortner},
  \bibinfo{title}{Recrystallization in hot working and creep},
  \bibinfo{journal}{Met. Sci.} \bibinfo{volume}{8}~(\bibinfo{number}{1})
  (\bibinfo{year}{1974}) \bibinfo{pages}{161{\textendash}167}.

\bibitem[{Sandstr{\"o}m and Lagneborg(1975)}]{sandstrom_model_1975-2}
\bibinfo{author}{R.~Sandstr{\"o}m}, \bibinfo{author}{R.~Lagneborg},
  \bibinfo{title}{A model for hot working occurring by recrystallization},
  \bibinfo{journal}{Acta Metall.} \bibinfo{volume}{23}~(\bibinfo{number}{3})
  (\bibinfo{year}{1975}) \bibinfo{pages}{387--398}, ISSN
  \bibinfo{issn}{0001-6160}, \doi{\bibinfo{doi}{10.1016/0001-6160(75)90132-7}}.

\bibitem[{Sandstr{\"o}m and Lagneborg(1974)}]{sandstrom_model_1974}
\bibinfo{author}{R.~Sandstr{\"o}m}, \bibinfo{author}{R.~Lagneborg},
  \bibinfo{title}{A model for hot working occuring by recrystallization},
  \bibinfo{journal}{Scripta Metallurgica}
  \bibinfo{volume}{8}~(\bibinfo{number}{11}) (\bibinfo{year}{1974})
  \bibinfo{pages}{liv--lv}, ISSN \bibinfo{issn}{0036-9748},
  \doi{\bibinfo{doi}{10.1016/0036-9748(74)90362-7}}.

\bibitem[{Ulam(1991)}]{ulam_adventures_1991}
\bibinfo{author}{S.~M. Ulam}, \bibinfo{title}{Adventures of a Mathematician},
  \bibinfo{publisher}{University of California Press}, ISBN
  \bibinfo{isbn}{9780520910553}, \bibinfo{year}{1991}.

\bibitem[{Von~Neumann(1951)}]{von_neumann_general_1951}
\bibinfo{author}{J.~Von~Neumann}, \bibinfo{title}{The general and logical
  theory of automata}, \bibinfo{journal}{Cerebral mechanisms in behavior}
  (\bibinfo{year}{1951}) \bibinfo{pages}{1{\textendash}41}.

\bibitem[{Hesselbarth and G{\"o}bel(1991)}]{hesselbarth_simulation_1991}
\bibinfo{author}{H.~Hesselbarth}, \bibinfo{author}{I.~G{\"o}bel},
  \bibinfo{title}{Simulation of recrystallization by cellular automata},
  \bibinfo{journal}{Acta Mater. et. Metall.}
  \bibinfo{volume}{39}~(\bibinfo{number}{9}) (\bibinfo{year}{1991})
  \bibinfo{pages}{2135--2143}, ISSN \bibinfo{issn}{0956-7151},
  \doi{\bibinfo{doi}{10.1016/0956-7151(91)90183-2}}.

\bibitem[{Ding and Guo(2001)}]{ding_coupled_2001}
\bibinfo{author}{R.~Ding}, \bibinfo{author}{Z.~X. Guo}, \bibinfo{title}{Coupled
  quantitative simulation of microstructural evolution and plastic flow during
  dynamic recrystallization}, \bibinfo{journal}{Acta materialia}
  \bibinfo{volume}{49}~(\bibinfo{number}{16}) (\bibinfo{year}{2001})
  \bibinfo{pages}{3163{\textendash}3175}.

\bibitem[{Kroc(2002)}]{kroc_application_2002}
\bibinfo{author}{J.~Kroc}, \bibinfo{title}{Application of cellular automata
  simulations to modelling of dynamic recrystallization}, in:
  \bibinfo{editor}{P.~M.~A. Sloot}, \bibinfo{editor}{A.~G. Hoekstra},
  \bibinfo{editor}{C.~J.~K. Tan}, \bibinfo{editor}{J.~J. Dongarra} (Eds.),
  \bibinfo{booktitle}{Computational Science {\textemdash} {ICCS} 2002}, no.
  \bibinfo{number}{2329} in \bibinfo{series}{Lecture Notes in Computer
  Science}, \bibinfo{publisher}{Springer Berlin Heidelberg}, ISBN
  \bibinfo{isbn}{978-3-540-43591-4, 978-3-540-46043-5},
  \bibinfo{pages}{773--782}, \bibinfo{year}{2002}.

\bibitem[{Kuc and Gawad(2011)}]{kuc_modelling_2011}
\bibinfo{author}{D.~Kuc}, \bibinfo{author}{J.~Gawad}, \bibinfo{title}{Modelling
  of microstructure changes in hot deformed materials using cellular automata},
  \bibinfo{journal}{{AIP} Conf. Proc.}
  \bibinfo{volume}{1315}~(\bibinfo{number}{1}) (\bibinfo{year}{2011})
  \bibinfo{pages}{1479--1484}, \doi{\bibinfo{doi}{10.1063/1.3552396}}.

\bibitem[{Kugler and Turk(2006)}]{kugler_study_2006}
\bibinfo{author}{G.~Kugler}, \bibinfo{author}{R.~Turk}, \bibinfo{title}{Study
  of the influence of initial microstructure topology on the kinetics of static
  recrystallization using a cellular automata model}, \bibinfo{journal}{Comp.
  Mater. Sci.} \bibinfo{volume}{37}~(\bibinfo{number}{3})
  (\bibinfo{year}{2006}) \bibinfo{pages}{284--291}, ISSN
  \bibinfo{issn}{0927-0256},
  \doi{\bibinfo{doi}{10.1016/j.commatsci.2005.07.005}}.

\bibitem[{Lee and Im(2010)}]{lee_cellular_2010}
\bibinfo{author}{H.~W. Lee}, \bibinfo{author}{Y.-T. Im},
  \bibinfo{title}{Cellular automata modeling of grain coarsening and refinement
  during the dynamic recrystallization of pure copper},
  \bibinfo{journal}{Materials Transactions}
  \bibinfo{volume}{51}~(\bibinfo{number}{9}) (\bibinfo{year}{2010})
  \bibinfo{pages}{1614}.

\bibitem[{Qian and Guo(2004)}]{qian_cellular_2004}
\bibinfo{author}{M.~Qian}, \bibinfo{author}{Z.~X. Guo},
  \bibinfo{title}{Cellular automata simulation of microstructural evolution
  during dynamic recrystallization of an {HY}-100 steel},
  \bibinfo{journal}{Mater. Sci. Eng., A}
  \bibinfo{volume}{365}~(\bibinfo{number}{1-2}) (\bibinfo{year}{2004})
  \bibinfo{pages}{180--185}, ISSN \bibinfo{issn}{0921-5093},
  \doi{\bibinfo{doi}{10.1016/j.msea.2003.09.025}}.

\bibitem[{Won~Lee and Im(2010)}]{won_lee_numerical_2010}
\bibinfo{author}{H.~Won~Lee}, \bibinfo{author}{Y.-T. Im},
  \bibinfo{title}{Numerical modeling of dynamic recrystallization during
  nonisothermal hot compression by cellular automata and finite element
  analysis}, \bibinfo{journal}{Int. J. Mech. Sci.}
  \bibinfo{volume}{52}~(\bibinfo{number}{10}) (\bibinfo{year}{2010})
  \bibinfo{pages}{1277--1289}, ISSN \bibinfo{issn}{0020-7403},
  \doi{\bibinfo{doi}{10.1016/j.ijmecsci.2010.06.003}}.

\bibitem[{Yazdipour et~al.(2008)Yazdipour, Davies, and
  Hodgson}]{yazdipour_microstructural_2008}
\bibinfo{author}{N.~Yazdipour}, \bibinfo{author}{C.~Davies},
  \bibinfo{author}{P.~Hodgson}, \bibinfo{title}{Microstructural modeling of
  dynamic recrystallization using irregular cellular automata},
  \bibinfo{journal}{Comp. Mater. Sci.}
  \bibinfo{volume}{44}~(\bibinfo{number}{2}) (\bibinfo{year}{2008})
  \bibinfo{pages}{566--576}, ISSN \bibinfo{issn}{0927-0256},
  \doi{\bibinfo{doi}{10.1016/j.commatsci.2008.04.027}}.

\bibitem[{Liu et~al.(1996)Liu, Baudin, and Penelle}]{liu_simulation_1996}
\bibinfo{author}{Y.~Liu}, \bibinfo{author}{T.~Baudin},
  \bibinfo{author}{R.~Penelle}, \bibinfo{title}{Simulation of normal grain
  growth by cellular automata}, \bibinfo{journal}{Scripta Mater.}
  \bibinfo{volume}{34}~(\bibinfo{number}{11}).

\bibitem[{Kugler and Turk(2004)}]{kugler_modeling_2004}
\bibinfo{author}{G.~Kugler}, \bibinfo{author}{R.~Turk},
  \bibinfo{title}{Modeling the dynamic recrystallization under multi-stage hot
  deformation}, \bibinfo{journal}{Acta Materialia}
  \bibinfo{volume}{52}~(\bibinfo{number}{15}) (\bibinfo{year}{2004})
  \bibinfo{pages}{4659--4668}, ISSN \bibinfo{issn}{1359-6454},
  \doi{\bibinfo{doi}{10.1016/j.actamat.2004.06.022}}.

\bibitem[{Sahni et~al.(1983)Sahni, Srolovitz, Grest, Anderson, and
  Safran}]{sahni_kinetics_1983}
\bibinfo{author}{P.~S. Sahni}, \bibinfo{author}{D.~J. Srolovitz},
  \bibinfo{author}{G.~S. Grest}, \bibinfo{author}{M.~P. Anderson},
  \bibinfo{author}{S.~A. Safran}, \bibinfo{title}{Kinetics of ordering in two
  dimensions. {II}. Quenched systems}, \bibinfo{journal}{Phys. Rev. B}
  \bibinfo{volume}{28}~(\bibinfo{number}{5}) (\bibinfo{year}{1983})
  \bibinfo{pages}{2705--2716}, \doi{\bibinfo{doi}{10.1103/PhysRevB.28.2705}}.

\bibitem[{Z{\"o}llner and Streitenberger(2004)}]{zollner_computer_2004}
\bibinfo{author}{D.~Z{\"o}llner}, \bibinfo{author}{P.~Streitenberger},
  \bibinfo{title}{Computer simulations and statistical theory of normal grain
  growth in two and three dimensions}, \bibinfo{journal}{Mater. Sci. Forum}
  \bibinfo{volume}{467-470} (\bibinfo{year}{2004}) \bibinfo{pages}{1129--1136},
  ISSN \bibinfo{issn}{1662-9752},
  \doi{\bibinfo{doi}{10.4028/www.scientific.net/MSF.467-470.1129}}.

\bibitem[{Cocks and Gill(1996)}]{cocks_variational_1996}
\bibinfo{author}{A.~Cocks}, \bibinfo{author}{S.~Gill}, \bibinfo{title}{A
  variational approach to two dimensional grain growth{\textemdash}I. Theory},
  \bibinfo{journal}{Acta Mater.} \bibinfo{volume}{44}~(\bibinfo{number}{12})
  (\bibinfo{year}{1996}) \bibinfo{pages}{4765--4775}, ISSN
  \bibinfo{issn}{1359-6454},
  \doi{\bibinfo{doi}{10.1016/S1359-6454(96)00121-8}}.

\bibitem[{Weygand et~al.(2004)Weygand, L{\'e}pinoux, and
  Br{\'e}chet}]{weygand_nucleation_2004}
\bibinfo{author}{D.~Weygand}, \bibinfo{author}{J.~L{\'e}pinoux},
  \bibinfo{author}{Y.~Br{\'e}chet}, \bibinfo{title}{On the Nucleation of
  Abnormal Grain Growth: A 2D Vertex Simulation}, \bibinfo{journal}{Mater. Sci.
  Forum} \bibinfo{volume}{467-470} (\bibinfo{year}{2004})
  \bibinfo{pages}{1123--1128}, ISSN \bibinfo{issn}{1662-9752},
  \doi{\bibinfo{doi}{10.4028/www.scientific.net/MSF.467-470.1123}},
  \bibinfo{note}{00000}.

\bibitem[{Lloyd(1982)}]{lloyd_least_1982}
\bibinfo{author}{S.~Lloyd}, \bibinfo{title}{Least squares quantization in
  {PCM}}, \bibinfo{journal}{Information Theory, {IEEE} Transactions on}
  \bibinfo{volume}{28}~(\bibinfo{number}{2}) (\bibinfo{year}{1982})
  \bibinfo{pages}{129{\textendash}137}.

\bibitem[{Nagai et~al.(1990)Nagai, Ohta, Kawasaki, and
  Okuzono}]{nagai_computer_1990}
\bibinfo{author}{T.~Nagai}, \bibinfo{author}{S.~Ohta},
  \bibinfo{author}{K.~Kawasaki}, \bibinfo{author}{T.~Okuzono},
  \bibinfo{title}{Computer simulation of cellular pattern growth in two and
  three dimensions}, \bibinfo{journal}{Phase Transitions}
  \bibinfo{volume}{28}~(\bibinfo{number}{1-4}) (\bibinfo{year}{1990})
  \bibinfo{pages}{177--211}, ISSN \bibinfo{issn}{0141-1594},
  \doi{\bibinfo{doi}{10.1080/01411599008207938}}.

\bibitem[{Gill and Cocks(1996)}]{gill_variational_1996}
\bibinfo{author}{S.~Gill}, \bibinfo{author}{A.~Cocks}, \bibinfo{title}{A
  variational approach to two dimensional grain growth{\textemdash}{II}.
  Numerical results}, \bibinfo{journal}{Acta Mater.}
  \bibinfo{volume}{44}~(\bibinfo{number}{12}) (\bibinfo{year}{1996})
  \bibinfo{pages}{4777--4789}, ISSN \bibinfo{issn}{1359-6454},
  \doi{\bibinfo{doi}{10.1016/S1359-6454(96)00122-X}}.

\bibitem[{Telley et~al.(1996)Telley, Liebling, and
  Mocellin}]{telley_laguerre_1996}
\bibinfo{author}{H.~Telley}, \bibinfo{author}{T.~M. Liebling},
  \bibinfo{author}{A.~Mocellin}, \bibinfo{title}{The Laguerre model of grain
  growth in two dimensions I. Cellular structures viewed as dynamical Laguerre
  tessellations}, \bibinfo{journal}{Philos. Mag. B}
  \bibinfo{volume}{73}~(\bibinfo{number}{3}) (\bibinfo{year}{1996})
  \bibinfo{pages}{395--408}, ISSN \bibinfo{issn}{1364-2812},
  \doi{\bibinfo{doi}{10.1080/13642819608239125}}.

\bibitem[{Sch{\"u}le(1996)}]{schule_justification_1996}
\bibinfo{author}{E.~Sch{\"u}le}, \bibinfo{title}{A justification of the Hillert
  distribution by spatial grain growth simulation performed by modifications of
  Laguerre tessellations}, \bibinfo{journal}{Comp. Mater. Sci.}
  \bibinfo{volume}{5}~(\bibinfo{number}{4}) (\bibinfo{year}{1996})
  \bibinfo{pages}{277--285}, ISSN \bibinfo{issn}{0927-0256},
  \doi{\bibinfo{doi}{10.1016/0927-0256(96)00004-3}}.

\end{thebibliography}

\end{document}